  \documentclass[12pt,preprint]{aastex}
  \usepackage{psfig}

  \slugcomment{Submitted May 7$^{th}$}

  \shorttitle{Minor Bodies with ACS}
  \shortauthors{Marchi et al.}

\begin{document}



  \title{Trails of Solar System Minor Bodies on WFC/ACS Images\altaffilmark{1}}

  \author{Simone Marchi\altaffilmark{2},
          Yazan Momany\altaffilmark{2},  and
          Luigi R. Bedin\altaffilmark{2}}

  \altaffiltext{1} {Based on observations  with the NASA/ESA {\it Hubble
  Space  Telescope} (GO9820),  obtained at  the Space  Telescope Science
  Institute, which  is operated by  AURA, Inc., under NASA  contract NAS
  5-26555.}   \altaffiltext{2}{Dipartimento di  Astronomia, Universit\`a
  di  Padova,   Vicolo  dell'Osservatorio  2,   I-35122  Padova,  Italy;
  marchi@pd.astro.it, momany@pd.astro.it, bedin@pd.astro.it}

  \begin{abstract} 

In this paper we analyse very short arcs of minor  bodies of the Solar
System detected  on Hubble Space  Telescope ($HST$) Wide Field Channel
ACS images.  In particular, we address how  to constrain the Keplerian
orbital elements  for  minor body detections,  illustrating the method
for  2 objects. One  of the  minor  bodies left 13 successive  trails,
making  it  the most well-sampled object  yet  identified in the $HST$
archive.  Most interestingly, we also address the problem of ephemeris
prediction and show  that in the particular case  of $HST$ very  short
arcs     the  confinement window  for   subsequent    recovery is {\em
significantly  reduced   to   a  narrow linear   region},   
that would facilitate successive observations.

  \end{abstract}


  %
  \section{Introduction}
  %

Although  the $HST$  was  not  specifically designed  to  serve as  an
asteroid surveyor, it  offers a unique opportunity to  study the solar
system minor objects (MOs). Indeed,  MO trails on the $HST$ images are
usually  characterized  by a  curved  shape,  whereas in  ground-based
observations they will  stand out only as straight  lines (for typical
exposure times).  The  curvature in $HST$ images is  simply due to the
parallax induced  by the telescope orbital motion  while exposing, and
is easily detectable owing to its high angular resolution. Taking into
account the  specific observation  conditions (i.e.\ the  ephemeris of
the  $HST$ while exposing),  it is  possible to  simulate trails  as a
function of geocentric distance and  line of sight. Thus, a comparison
of  the observed  trail with  the simulated  ones, enables  a distance
determination of the detected MOs.

This method has been illustrated in  the work of Evans et al.\ (1998),
which presented trails of 96  MOs, detected on 28,460 WFPC2 deep $HST$
images.  Most  of these  MOs are  too faint to  have been  detected in
ground-based observations.   In this regard, the recent  advent of the
Advanced Camera  for Surveys  (ACS) offers new  perspectives.  Indeed,
the Wide Field Channel (WFC) of the ACS has a smaller pixel scale than
the  wide field  cameras of  WFPC2;  it is  up to  $\sim$5 times  more
sensitive; and most interestingly, it  offers a field of view which is
two times wider.  These aspects  are of significant impact in a better
tracking of the trails, lowering  the limit of detectable objects; and
increasing the probability of  serendipitous capture of MOs within the
WFC/ACS field  of view.  Extrapolating  from the Evans et  al.\ (1998)
results,  one finds that,  on average,  one in  every thirty  ACS deep
exposures may unveil the trail of a new faint MO.

Once a MO has been detected,  the next step would be the determination
of an orbital solution, in  order to calculate accurate ephemeris. The
knowledge of these two is indispensable for any further investigation.
The problem of MOs orbital  determination has been widely discussed in
the  literature.   Well  defined  orbital  solutions  require  several
observations  spanning  periods  of  time  long  enough  to  show  the
curvature of their  motion.  However, in most cases  the observed arcs
are too short to obtain convergent solutions (the so-called very short
arcs,  VSAs, see  Milani et  al.\  2004).  In  general, MO  detections
originate  from: (1) automated  surveys, e.g.   LONEOS or  LINEAR; (2)
casually  detected  objects,  e.g.\  observations obtained  for  other
scientific purposes (as in our  case), or (3) digital archive surveys,
e.g.  Barbieri et al.\ (2004).   In all these cases, determinations of
orbital solutions and geocentric distances from VSAs are not possible.

In  this paper we  deal with a particular  kind of  VSAs, namely those
detected with  the  Hubble  Space Telescope  (HVSAs).   HVSAs can   be
considered  an  intermediate   case between   ground-based   VSAs  and
detections for which full orbital solutions  are determined.  The main
difference derives   from the fact that  for   HVSAs one can determine
geocentric distances even if the  arcs are very  short (e.g.  1-2 h or
less).  This  has a considerable impact  on constraining the orbital
solutions.  Although a unique solution  is still out of reach,  useful
information  concerning the  real  nature   of the  detection can   be
obtained.  Moreover, the knowledge of geocentric distances is vital in
recovering these objects in successive  observations. We discuss these
topics by providing two examples of MOs detected on $HST$/ACS images.

  %
  \section{Observations and Data Reduction}
  %

The  $HST$/ACS data  come from  GO-9820 (P.I.:  Momany), aimed  at the
study of  the star  formation history of  the Local  Group Sagittarius
dwarf irregular  galaxy (SagDIG).  The central pointing  of all images
were  centered   at:  ($\alpha$=19:29:58.97,  $\delta$=$-$17:40:41.3),
whose   ecliptic  coordinates  are:   $\lambda$=291.437$^{\circ}$  and
$\beta$=+4.103$^{\circ}$ at  J2000 equinox.  The  observations consist
of  a total  number of  3 $HST$  orbits, one  orbit per  filter  and 5
exposures per orbit: 5$\times$396s in F606W ($V_{606}$), 5$\times$419s
in F475W ($B_{475}$), and 5$\times$419s in F814W ($I_{814}$),
conducted on August 18$^{th}$ 2003, covering a  total time interval of
$\sim3.9$  hours.  These  images  were preprocessed with the  standard
STScI pipeline. A  correction of the  geometrical distortion  has been
applied following the Anderson (2002) recipe.  

  \section{The Two MOs}

  \subsection{Identification}

\begin{figure*}
 \centerline{\psfig{figure=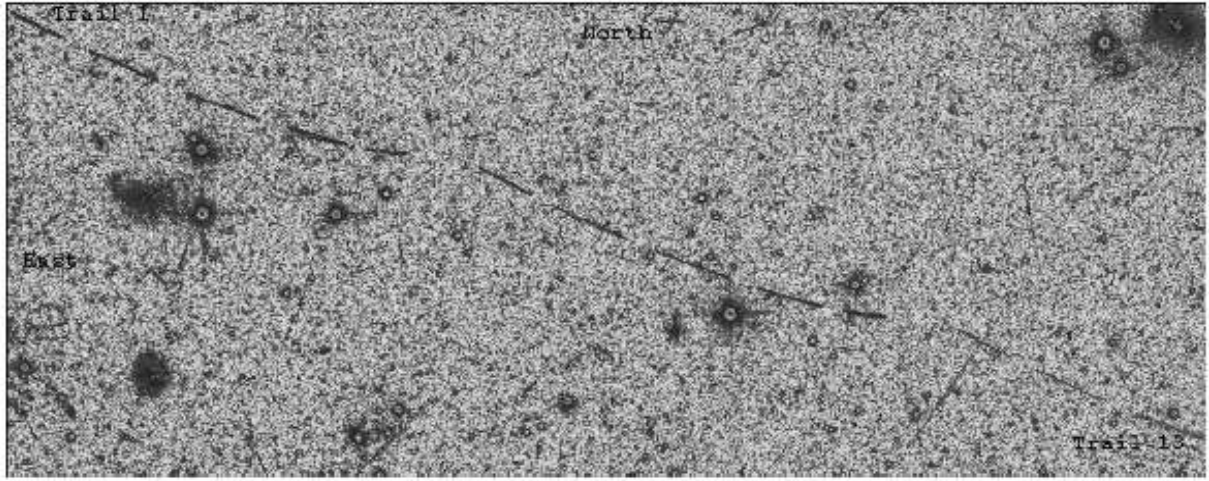,width=16cm}}
  \caption{The 13 trails  left by   MO1  on  the sum image.  North
  is up, East is to the left.  The first trail is the upper left one: MO1
  at $\alpha$=19:29:56.602 and  $\delta$=$-$17:40:00.01. The last trail is
  the  lower right  one,  having crossed  $\Delta \alpha=-65.1$  $\Delta
  \delta=-25.43$ arcsec. The first 5 trails are in $V_{606}$ filter,
  followed by 5 trails in $B_{475}$ and 3 trails in $I_{814}$.}
  \label{trail}
  \end{figure*}

A rapid check  for passing MOs has been done by  eye inspection of all
available images.  Using a selected sample of common stars, all single
images were put onto a common reference system.  The image combination
was simply  a sum, i.e.  all  particular features like  cosmic rays or
transiting MOs were preserved.  Although this method resulted in a sum
image  that is  full of  cosmic  rays, on  the other  hand, trails  of
transiting  MOs  stood-out as  these  formed  a  sequence of  multiple
trails, that were easily detected.  This method helped to identify two
MOs with multiple trails.
Prior to  this study,  the two  MOs were not  registered in  the Minor
Planet                Center                (MPC)                ({\sf
http://scully.harvard.edu/$^{\sim}$cgi/CheckMP}) database.
The brighter  one (MO1  hereafter), left 13  consecutive curved-trails
covering  3 $HST$  orbits (5  trails  in the  $V_{606}$ and  $B_{475}$
filters  and 3  trails in  the $I_{814}$  filter).  The  second object
(MO2) left only 3 trails, all of them in the $V_{606}$ filter.  Figure
\ref{trail} shows the 13 trails of MO1 on the sum image.  It is by far
the  most well-sampled  asteroid  trail yet  identified  in the  $HST$
archive data.  MOs with even one trail can still be distinguished from
cosmic rays, as MOs will  reflect the telescope point spread function.
We however could not identify additional objects.
%


  %
  \subsection{Distances and Orbits}
  %

\begin{figure*}
  \centerline{\psfig{figure=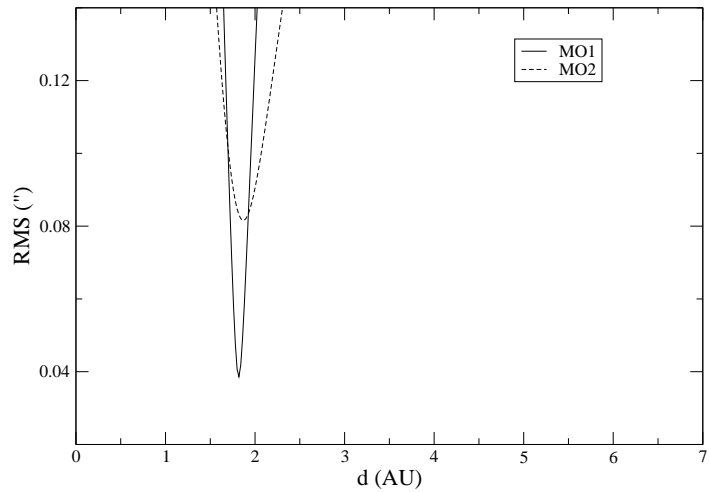,width=10cm,angle=-90}}
  \caption{Root mean square (RMS) differences (in arcsec) between 
the start/end points of the observed trails and simulated  trajectories 
(see text).}
  \label{result_fit}
  \end{figure*}

Our analysis of  individual trails follows  the guidelines of the work
of Evans  et al.\  (1998).  In general,  the apparent motion of a minor
body in the sky is  due to the  superposition of the observer's motion
and the intrinsic motion of the object. For  objects identified in $HST$
images, the apparent motion $\vec{P}$ can be written:

  $$ \vec{P}(t) = \vec{P}_{\rm H}(t)+\vec{P}_{\rm E}(t)+\vec{P}_{\rm MO}(t) $$
where $\vec{P}_{\rm H}, \vec{P}_{\rm E}, \vec{P}_{\rm MO}$  
are the parallaxes due  to
$HST$, the Earth, and the intrinsic minor body motion, respectively.  By
knowing $\vec{P}_{\rm H},\vec{P}_{\rm E}$,  
and measuring $\vec{P}$ on the images,
one can  determine the  intrinsic motion  rate of  the  minor body  as
follows:

$$ \vec{P}_{\rm MO}(t,d(t)) = \vec{P}(t)-\vec{P}_{\rm E}(t,d(t))-\vec{P}_{\rm H}(t,d(t)) $$
where we have now explicitly indicated the dependence of $\vec{P}_{\rm
H}, \vec{P}_{\rm  E}$, and  hence of $\vec{P}_{\rm  MO}$, on  the MO's
geocentric distance  $d$, which also varies with  time.  The intrinsic
rate  of the  minor  body does  change  with time.  However, on  short
timescales  ($\sim$2-3  hours),  we   assume  that  it  has  a  linear
dependence  with time, i.e.   $\vec{P}_{\rm MO}(t,d(t))$=$\vec{P}_{\rm
MO}(d)\times t$, and that $d$ is  constant.  So, for each value of $d$
we compute the corresponding $\vec{P}_{\rm MO}$ using the maximum time
interval available, i.e.  endpoints  of the whole exposure sequence in
the same filter.  This allows  us to reconstruct the trajectory of the
body, for any value of $d$ and  at any given time.  On the other hand,
the best value of $d$ is determined by minimizing the root mean square
differences between the start/end  points of the {\it observed} trails
and {\it simulated} trajectories (see Fig.~\ref{result_fit}).
For   this task, only the  start/end  points of {\em  each} trail were
used, as these are the only points with time tags.
Although, the actual  shape  of each trail  has  not been  used in the
fitting  procedure [in this aspect our   analysis differs from that of
Evans  et  al.\ (1998)], it   remains  that our   best estimate of  $d$
reproduces the shape of {\em all observed trails}.
The resulting geocentric  (heliocentric) distance for  MO1 and MO2 are
$d$=$1.82  (2.72)  \pm0.10$~AU and      $d$=$1.86  (2.76) \pm0.15$~AU,
respectively. The larger error  for MO2 is due to the smaller number
of measured points.

\begin{figure*}
  \centerline{\psfig{figure=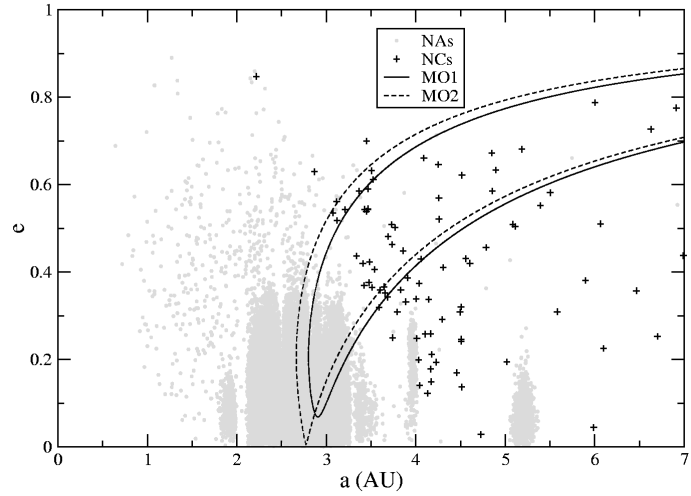,width=10cm,angle=-90}}
  \caption{Orbital solutions for the detected objects  in the $a,e$ plane (see
 text). The $a$ axis has been terminated at 7 AU in order to better show
 the region of the Main Belt and Jupiter Trojans.}
  \label{result_a_e}
  \end{figure*}

\begin{figure*}
  \centerline{\psfig{figure=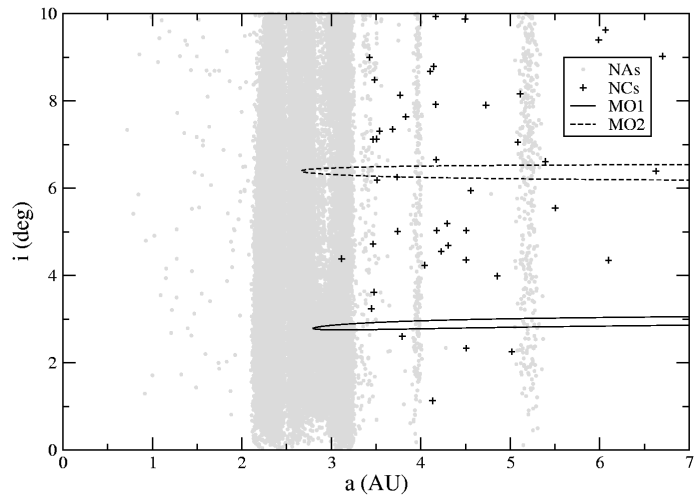,width=10cm,angle=-90}}
  \caption{Orbital solutions for the detected objects  in the  $a,i$ plane
 (see text and Fig. \ref{result_a_e}).}
  \label{result_a_i}
  \end{figure*}

Once the geocentric distance of  the minor body is determined, one can
compute its geocentric velocity in the plane perpendicular to the line
of sight.   However, the velocity  component along the line  of sight,
$\dot{d}$, cannot be determined.  For  this reason it is impossible to
obtain a single orbital solution, as it would require the knowledge of
$d$ and $\dot{d}$.  However, it is possible to put some constraints on
the  orbital  elements.   In  order  to have  a  bound  solution,  the
heliocentric velocity  has to  be such that  the total  energy, ${\cal
E}(d,\dot{d})$, is $<0$.  Hence, for each value of $\dot{d}$ for which
${\cal E}(d,\dot{d})<0$ is satisfied, a corresponding orbital solution
is obtained.  The  eccentricity in these solutions can  vary from 0 to
1,  leading   to  a   large  variation  of   $a$  values.    In  Figs.
\ref{result_a_e} and \ref{result_a_i} the Keplerian elements $a, e, i$
for  the possible solutions  are plotted.  It is  seen that  while the
solutions for  $a$ and  $e$ are not  well determined,  the inclination
instead is  restricted to  a narrow region.   Figures \ref{result_a_e}
and \ref{result_a_i} also show double solutions for each value of $a$,
corresponding to solutions with  $\dot{d}$ being positive or negative.
On the same plots, numbered  asteroids (NAs) and numbered comets (NCs)
distributions are shown for comparison.  Judging the range of possible
values for $a$,  $e$ and $i$, it  is most likely that our  two MOs are
main belt asteroids, although other solutions cannot be ruled out.

 \subsection{HVSA ephemeris}

\begin{figure*}         
  \centerline{\psfig{figure=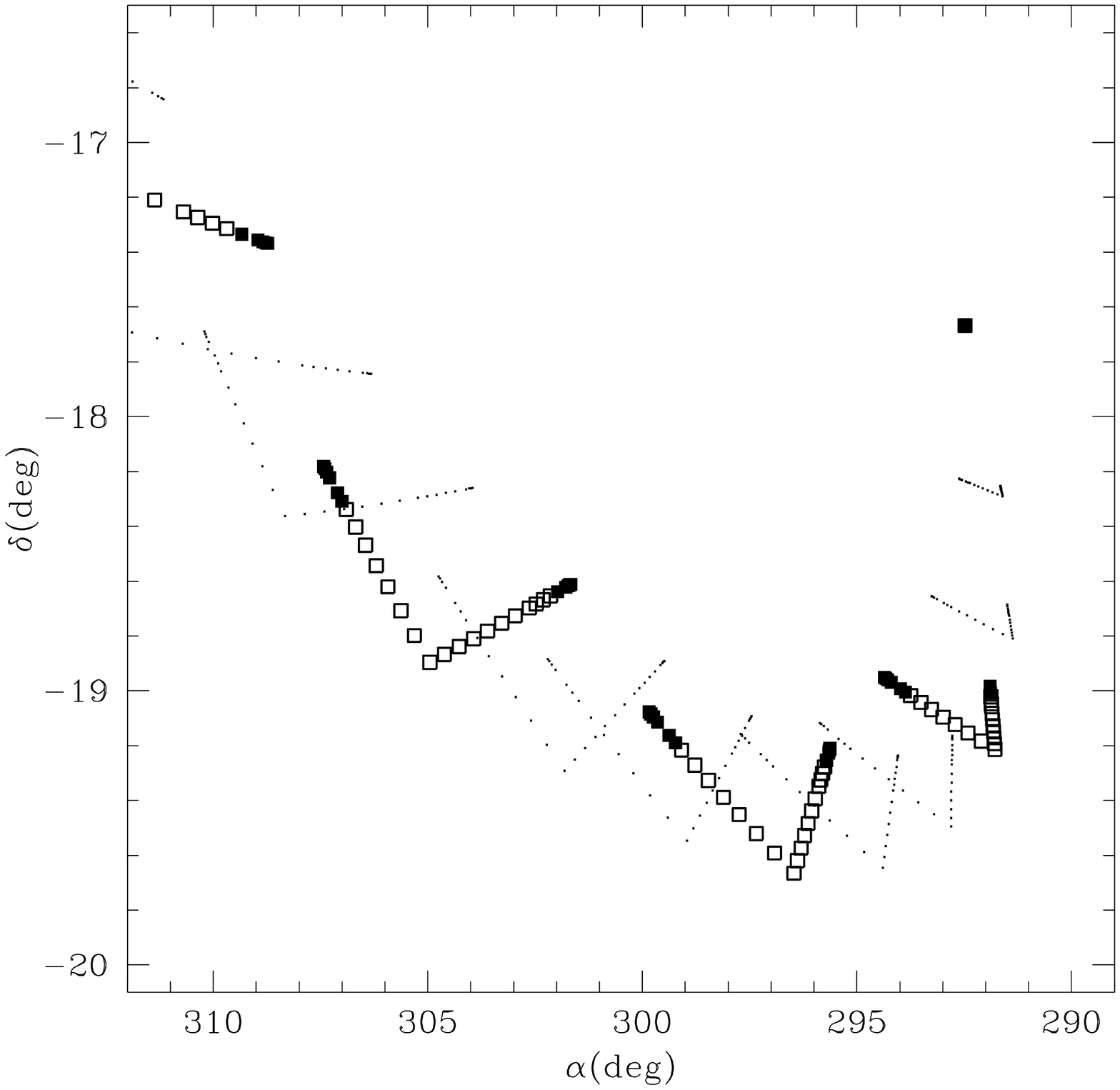,width=14cm}}
  \caption{Linear ephemeris for MO2. The plotted points span a 10 days
  interval,  while  squared symbols  span  a 30   days interval.  The
  points  (29 in   total for    each  date)  sample the solution    of
  Figs.~\ref{result_a_e}  and \ref{result_a_i},   where only $a<100$AU
  have   been  considered.  Open  squares  highlight  solutions having
  $a<7$~AU  (the most probable ones), while  filled ones are for those
having  $7<a<100$~AU. The single point   on the upper  right corner   is the
  starting date for which    all  the solutions  are coincident.
    The typical  ``V'' shape of  the  distribution corresponds to the double
  solutions (see text) in Figs.~\ref{result_a_e} and
\ref{result_a_i}, while  the vertices  correspond  to solutions  about
$e$  minimum.   The linear regions  follow each  other in  a clockwise
direction reflecting mostly the Earth's motion.}

 \label{ephem}
  \end{figure*}

Although the observed arcs are very  short, thanks to the fact that we
are  dealing  with HVSAs  the  possible  orbital  solutions have  been
significantly  constrained. We now  examine whether  another important
aspect of  MO detections can be  constrained, that is  the recovery of
these objects in successive observations.
The problem of generating  ephemeris for VSAs detected by ground-based
telescopes has been  discussed already by Milani et  al.\ (2004).  The
main  result of  their analysis  is the  production  of ``triangulated
ephemeris'', i.e.  ephemeris computed on  a grid of  orbital solutions
which sample  in a proper way  the VSA admissible region  in the plane
$(d,\dot{d})$.
This ephemeris production for each point  of the grid results in a two
dimensional portion of the sky where the MO has to be confined.
The size of this   confinement region increases with time.  Therefore,
and to be useful for an actual recovery, this  ``window'' has to be as
narrow as possible.
This is particularly important when dealing with HVSAs. Indeed, thanks
to  the knowledge of  the geocentric  distance $d$,  we find  that the
admissible region becomes  a {\it line} instead of  an {\it area}, and
consequently  the  triangulated ephemeris  becomes  a ``linear''  one.
This situation  is illustrated in  Fig.~\ref{ephem}.  This drastically
reduces the number of frames required to sample the permitted area for
the recovery of the objects. Figure~\ref{ephem} shows that mapping the
linear  ephemeris would still  require a  reasonable number  of frames
(depending  on the  telescope used)  even 2-3  months after  the first
detection.

  \subsection{Photometry}

For  each trail,  aperture  photometry is  performed  on every  single
image.    Since   the   images   were  preprocessed   (de-biased   and
flat-fielded) by  the standard STScI pipeline, we  directly summed the
counts  within 0.5  arcsec of  each pixel  in the  trail. A  local sky
estimate  was obtained  in an  adjacent  area after  the rejection  of
cosmic rays.  The obtained digital  counts were then put onto the Vega
System magnitude following the recipe  in Holtzman et al.\ (1995), and
in the particular case of ACS observations as in Bedin et al.\ (2004).
Table~\ref{mag}  reports the derived magnitudes for  MO1  and MO2, and
allows us to address possible  variations of their magnitude with time
(possibly related to rotation).
Given the errors (of the order of 0.04 and 0.07 magnitudes for MO1 and
MO2) there is no clear indication of variability for MO1. On the other
hand, the  $V_{606}$ magnitude  of MO2 varies  considerably, although,
with only 3 exposures no firm conclusion can be reached.

MO1 has been detected in many frames, and most interestingly, in three
different filters ($V_{606}$, $B_{475}$, $I_{814}$). The color indices
of MO1  may shed light on its  nature.  For this task  we computed the
solar  color   indices  in  the  same  photometric   system,  that  is
$B^{\odot}_{475}-V^{\odot}_{606}$=$0.570$                           and
$V^{\odot}_{606}-I^{\odot}_{814}$=$0.569$.   The  derived reflectivity
(defined              as              $R_i$=$10^{-0.4((i-V_{606})_{\rm
MO}-(i-V_{606})_{\odot})}$),  is:  $R_{475}$=$1.03$,  $R_{606}$=$1.00$
and $R_{814}$=$1.20$.   MO1's reflectivity indicates a  red surface, a
common property among outer MOs, like X-, M- D-type objects.  Finally,
assuming  a geometric  albedo  of  0.1 in  the  $V_{606}$ filter,  the
estimated sizes of MO1 and MO2 are about 2.4 and 0.7 km, respectively.
This  further  confirms  the  capability  of $HST$  to  observe  small
objects.

  \begin{table*}[h]
  {\footnotesize
  \caption{Magnitudes of the two observed minor bodies.}
  \label{mag}
  \begin{center}
  \def\V{\rule{0pt}{3.5ex}}
  \begin{tabular}{l|ccc}
  \hline \hline \hline
  \V \bf Filter & \bf MO1    & \bf MO2     \\[1.3ex]
  \hline
  \V $B_{475}$\#1   &  20.46         &     \\
     $B_{475}$\#2   &  20.41         &     \\
     $B_{475}$\#3   &  20.30         &     \\
     $B_{475}$\#4   &  20.34         &     \\
     $B_{475}$\#5   &  20.34         &     \\ [1.3ex]
  \hline
  \V $V_{606}$\#1   &  19.78         & 22.38    \\
     $V_{606}$\#2   &  19.77         & 22.28    \\
     $V_{606}$\#3   &  19.76         & 22.00    \\
     $V_{606}$\#4   &  19.92         &         \\
     $V_{606}$\#5   &  20.00         &     \\ [1.3ex]
  \hline
  \V $I_{814}$\#1   &  19.06         &     \\
     $I_{814}$\#2   &  19.05         &     \\
     $I_{814}$\#3   &  19.12         &     \\ [1.3ex]
  \hline \hline \hline
  \end{tabular}
  \end{center}
  }
  \end{table*}


  %
  \section{Conclusions}
  %

In this paper we report the  identification of very short arcs (from 2
minor  bodies) detected  in $HST$  WFC/ACS images,  and  constrain the
Keplerian elements  of their  orbits.  Most interestingly,  we address
the problem  of ephemeris prediction  and show that in  the particular
case of  HVSAs the confinement  window for subsequent recovery  of the
MOs is  significantly reduced  to {\em a  narrow linear  region}, that
would facilitate subsequent observations.

 \acknowledgments

We thank  the anonymous referee  for interesting comments  that helped
improve the layout of the paper. We also thank Prof.\ C.\ Barbieri for
useful discussions and  Dr.\ M.\ Clemens for a  careful reading of the
manuscript.

  \end{document}